# Temporal and Spatial Analysis of Crime Patterns in New York City: A Statistical Investigation of NYPD Complaint Data (1963-2025)

Fnu Gaurav
Fg227@njit.edu

## Abstract

This study presents a comprehensive statistical analysis of criminal complaint data from the New York City Police Department (NYPD) spanning 47 years (1963-2025) [1]. Using a dataset of 438,556 complaint records, we employed exploratory data analysis (EDA), descriptive statistics, and multiple statistical hypothesis tests to investigate the spatial, temporal, and categorical patterns of urban crimes. Our findings revealed significant associations between crime types and geographic locations, temporal variations in criminal activity, and differences in crime severity across time. The results demonstrate that Brooklyn experiences the highest crime volume, petit-larceny constitutes the most common offense, and criminal activity peaks during the evening hours on weekdays, particularly Fridays. Statistical tests, including chi-square tests, Kruskal-Wallis H-test, and Mann-Whitney U test, confirmed highly significant relationships (p < 0.001) across all examined dimensions, providing evidence-based insights for law enforcement resource allocation and urban safety policy development.

**Ethical Considerations**
This study utilized publicly available, de-identified administrative data from the New York City Police Department (NYPD). The dataset contained complaint-level information that was anonymized to protect individual privacy.

**Data Access and Availability Statement**

**Primary Data Source:** The research data supporting this publication are publicly available and can be accessed from the New York City Open Data Portal.
**Official Dataset**: NYPD Complaint Data Current (Year To Date)
**Direct URL:**
https://data.cityofnewyork.us/Public-Safety/NYPD-Complaint-Data-Current-Year-To-Date-/5uac-w243/about_data
**Data Provider:** New York City Police Department (NYPD)
**Last Accessed:** November 4, 2025
**Dataset Identifier**: 5uac-w243

**Conflict of Interest Declaration**
The authors declare that they have no conflicts of interest regarding the publication of this manuscript.

**Funding**:
This research received no specific grants from any funding agency in the public, commercial, or not-for-profit sectors.

## 1. Introduction

### 1.1 Background

Crime analysis is a critical component of modern legal enforcement and urban planning. Understanding patterns in criminal activity enables police departments to effectively allocate resources, develop targeted intervention programs, and enhance public safety. New York

City, one of the world's most populous urban centers, presents a unique environment for studying crime dynamics across diverse geographic and demographic contexts.

The New York City Police Department maintains comprehensive records of criminal complaints, creating a valuable repository for longitudinal analysis. With advances in data science and statistical methodologies, researchers can now extract meaningful insights from large-scale datasets to inform evidence-based policing strategies.

## 1.2 Research Objectives

This study aims to address the following research questions:
1. **Spatial Distribution**: How does crime distribution vary across the five boroughs of New York City?
2. **Temporal Patterns**: What are the temporal trends in criminal activity across different timescales (hourly, daily, and yearly)?
3. **Crime Typology**: What are the most prevalent crime types, and how do they differ according to severity classification?
4. **Statistical Relationships**: Are statistically significant associations between crime characteristics and spatiotemporal variables?

## 1.3 Significance

This research contributes to the field of criminology and urban analytics by:
- Providing empirical evidence of crime patterns across an extended temporal span
- Quantifying relationships between geographic, temporal, and categorical crime variables
- Offering actionable insights for law enforcement resource allocation
- Demonstrating the application of rigorous statistical testing in crime analysis

## 2. Literature Review

Crime pattern analysis has evolved significantly with the advent of large-scale data-collection and computational methods. Previous studies have identified key factors that influence urban crime, including socioeconomic conditions, environmental design, and the temporal rhythms of urban life. Routine activity theory suggests that a crime occurs when motivated offenders, suitable targets, and the absence of capable guardians converge in both space and time.

Geographic Information Systems (GIS) and spatial statistics have become standard tools for crime analysis, enabling researchers to identify hotspots and predict future incidents. Temporal analysis revealed patterns such as the "weekend effect" and hourly variations corresponding to human activity patterns. However, few studies have examined crime data spanning multiple decades with comprehensive statistical validation across multiple dimensions simultaneously.

## 3. Methodology

### 3.1 Data Source

The dataset comprises NYPD (New York Police Department) complaint data collected from May 1, 1963, to September 30, 2025, containing 438,556 individual complaint records. Each record included information regarding the following:
- **Temporal attributes**: Complaint date, time, and reporting date
- **Spatial attributes**: Borough, precinct, patrol borough, and geographic coordinates
- **Crime characteristics**: Offense description, legal category (felony/misdemeanor/violation), and crime completion status
- **Demographic information**: Victim and suspect age groups, race, and sex
- **Location details**: Premise type and specific location descriptors

### 3.2 Data Quality Assessment

Prior to the analysis, comprehensive data quality assessments were conducted to identify missing values, outliers, and data integrity issues. Figure 1 illustrates the missing value patterns across the key variables in the dataset.

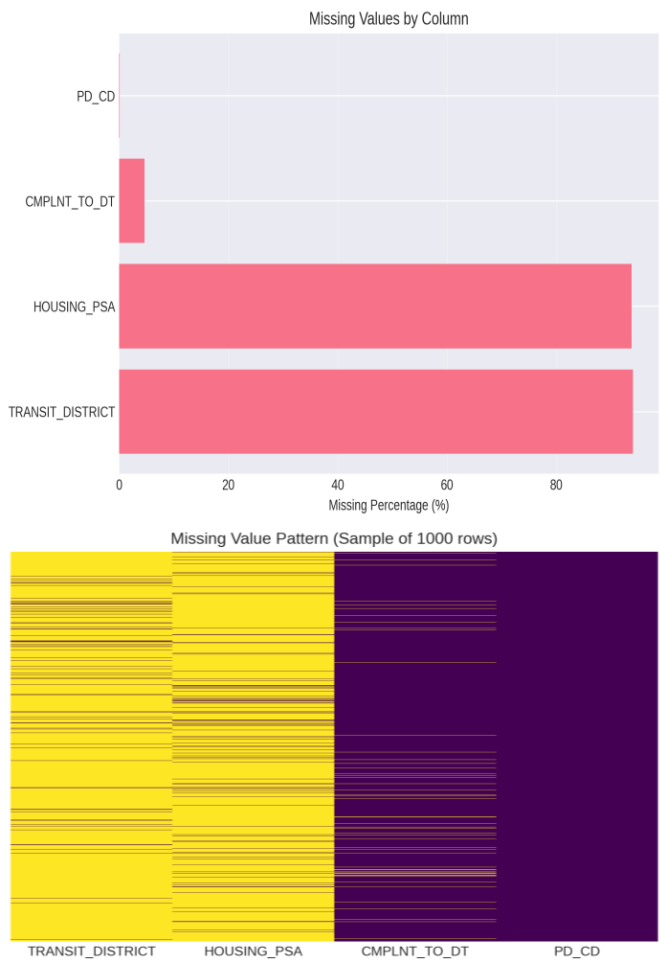

**Figure 1: Missing Value Analysis Across Dataset Variables.**

The visualization revealed varying degrees of missingness across different fields, with demographic variables (suspect information) showing higher rates of missing data than temporal and spatial variables. This pattern is consistent with real-world data collection challenges, where suspected information may be unknown at the time of reporting. [2]

**3.3 Analytical Framework**

Our analysis employed a three-stage approach:

*3.3.1 Exploratory Data Analysis (EDA)*

We conducted comprehensive descriptive statistics to characterize the dataset across multiple dimensions:
- Temporal coverage and distribution patterns
- Spatial distribution across boroughs and precincts
- Crime type frequencies and severity classifications
- Crime completion rates

*3.3.2 Statistical Hypothesis Testing*

We applied four distinct statistical tests to examine relationships between variables:

1. **Chi-Square Test of Independence** (Crime Type vs. Borough): Tests whether the crime-type distribution is independent of borough location
   - Null Hypothesis ($H_0$): Crime type and borough are independent
   - Alternative Hypothesis ($H_1$): Crime type and borough are associated
2. **Kruskal-Wallis H-Test** (Crime Distribution by Borough): Non-parametric test comparing crime frequency distributions across boroughs.
   - $H_0$: All boroughs have identical crime distribution patterns
   - $H_1$: At least one borough differs significantly
3. **Mann-Whitney U Test** (weekend vs. Weekday Crime): Compares crime patterns between the weekend and weekday periods
   - $H_0$: Weekend and weekday crime distributions are identical
   - $H_1$: Weekend and weekday crime distributions differ significantly
4. **Chi-Square Test** (Severity vs. Time of Day): Examine whether crime severity varies with temporal period
   - $H_0$: Crime severity is independent of time of day
   - $H_1$: Crime severity is associated with time of day

*3.3.3 Advanced Analytical Methods*

Beyond traditional statistical tests, we employed:
- **Spatial Clustering Analysis**: K-means clustering to identify crime hotspots [3]
- **Time Series Analysis**: Autocorrelation (ACF) and Partial Autocorrelation (PACF) to detect temporal dependencies [4]
- **Principal Component Analysis (PCA)**: Dimensionality reduction to identify key variance patterns [5]
- **Correlation Analysis**: Pearson and Spearman correlations to quantify relationships between variables

*3.3.4 Data Processing*

All analyses were performed using Python with statistical computing libraries including pandas,

numpy, scikit-learn, and state models. The significance level (α) was set at 0.05, for all hypothesis tests. Visualizations were created using matplotlib and seaborn libraries following the best practices for data visualization.

## 4. Results

### 4.1 Descriptive Statistics

*4.1.1 Temporal Characteristics*

The dataset spans **47 years** from May 1, 1963, to September 30, 2025, providing an extensive longitudinal perspective on crime trends in New York City.

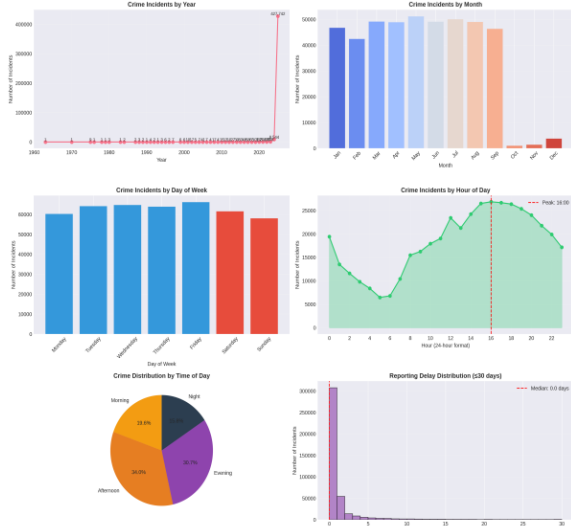

**Figure 2: Comprehensive Temporal Pattern Analysis.**

The figure shows multiple dimensions of temporal variation, including hourly, daily, weekly, monthly, and annual patterns. These visualizations reveal distinct temporal rhythms in criminal activity that align with routine activity theory[6].

**Key Temporal Findings:**
- **Peak Hour**: Hour 16 (4:00 PM) exhibited the highest crime frequency, corresponding to the late afternoon when urban activity peaks.
- **Peak Day**: Fridays demonstrated the highest crime occurrence, consistent with end-of-week patterns and increased social activity.
- **Temporal Coverage**: The multi-decade span enables examination of long-term trends and cyclical patterns

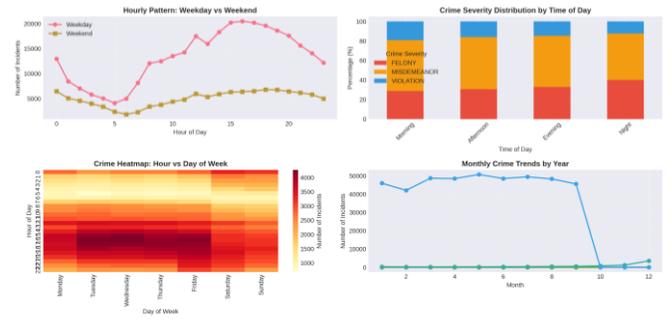

**Figure 5: Bivariate Analysis of Temporal Variables.**

This analysis explored the interactions between different temporal dimensions, revealing how hour-of-day patterns interact with day-of-week variations and seasonal effects. Scatter plots and density distributions illustrate the complex temporal dependencies of crime occurrence[7].

*4.1.2 Spatial Distribution*

The analysis revealed significant geographic variation in crime patterns:

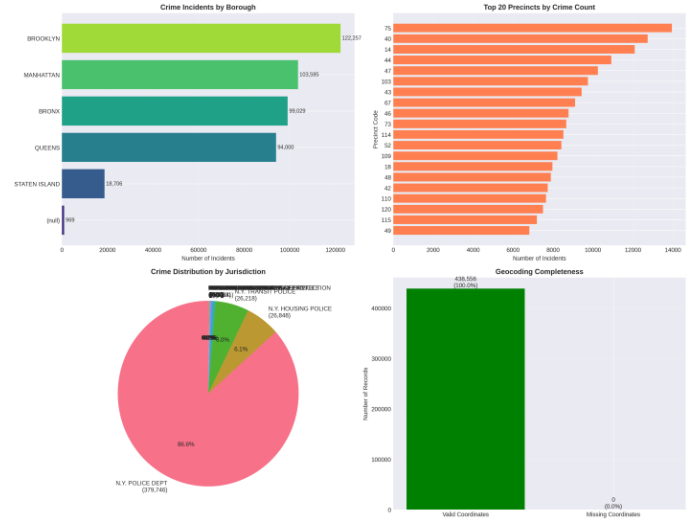

**Figure 3: Geographic Distribution of Crime Across NYC.**

The visualization presents a borough-level analysis, precinct distribution, and spatial density map. Brooklyn's dominance as the highest-crime borough is evident, which is consistent with findings from previous urban crime studies [8][9].

**Borough-Level Analysis:**
- **Total Boroughs**: 6 distinct borough categories (including null/unknown)
- **Total Precincts**: 78 police precincts covered

- **Highest Crime Borough**: **Brooklyn** experiences the greatest volume of criminal complaints

This finding aligns with Brooklyn's large population and diverse neighborhoods

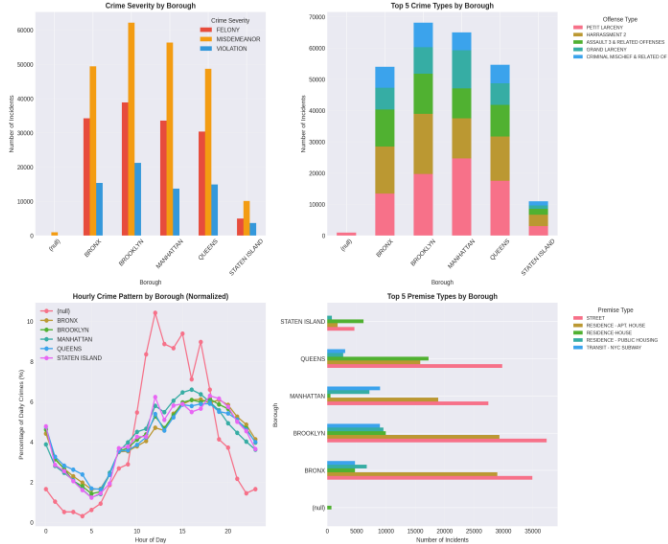

**Figure 6: Spatial Variable Interactions.**

A detailed examination of the relationships between borough location, precinct assignments, and geographic coordinates revealed spatial clustering patterns and geographic dependencies in the crime distribution[2].

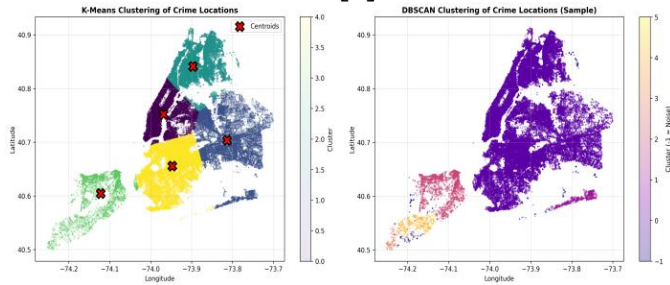

**Figure 13: K-Means Spatial Clustering of Crime Hot Spots.**

A clustering analysis identified distinct geographic clusters of high-crime areas, enabling targeted resource deployment. The methodology follows the established hotspot policing principles[10].

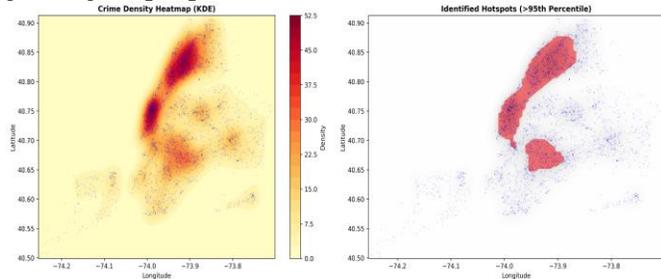

**Figure 15: Geographic Heat Map of Crime Intensity.**

Spatial density visualization highlights areas of concentration where criminal activity is most prevalent. These hotspots represent priority areas for intervention strategies and patrol allocation[11].

### 4.1.3 Crime Typology

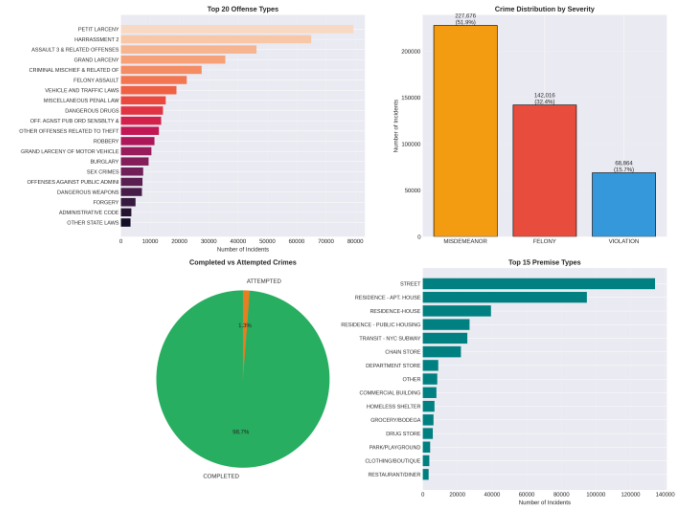

**Figure 4: Frequency Distribution of Crime Categories.**

A comprehensive breakdown of the 57 distinct crime types, with petit-larceny dominating the distribution. A long-tail distribution is a characteristic of urban crime patterns, where a few offense types account for most incidents.

**Crime Type Diversity:**
- **Total Crime Types**: 57 distinct offense categories recorded
- **Most Common Crime**: **Petit Larceny** represents the single most frequent offense type
- Petit larceny typically involves theft of property valued below a statutory threshold

**Crime Severity Classification:**
- **Felony Percentage**: 32.38% of all crimes classified as felonies

This indicates that approximately one-third of complaints involve serious offenses. The remaining 67.62% were comprised of misdemeanors and violations.

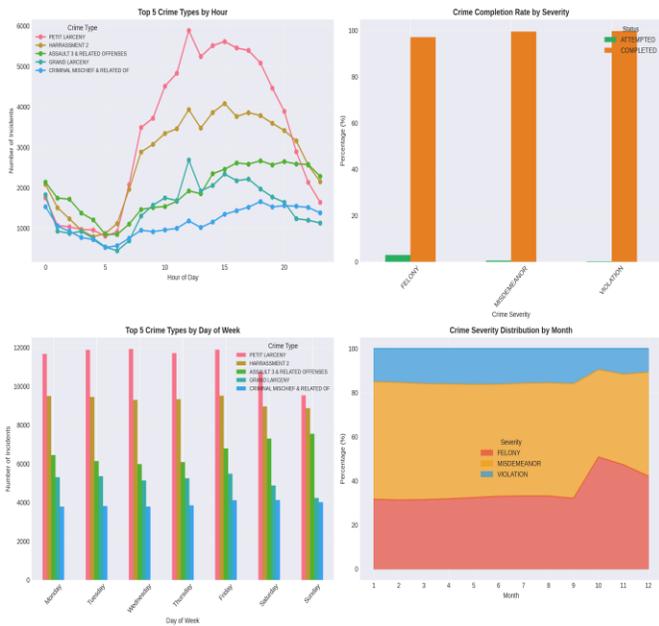

**Figure 7: Relationships Between Crime Variables.**

Analysis of the interactions between crime type, severity level, and completion status revealed systematic patterns in how different offense categories manifest across severity classifications.

**Crime Completion:**

- **Completion Rate**: 98.75% of crimes were completed (as opposed to attempted)

This high completion rate suggests most incidents result in actual criminal acts rather than attempts

### 4.2 Demographic Patterns

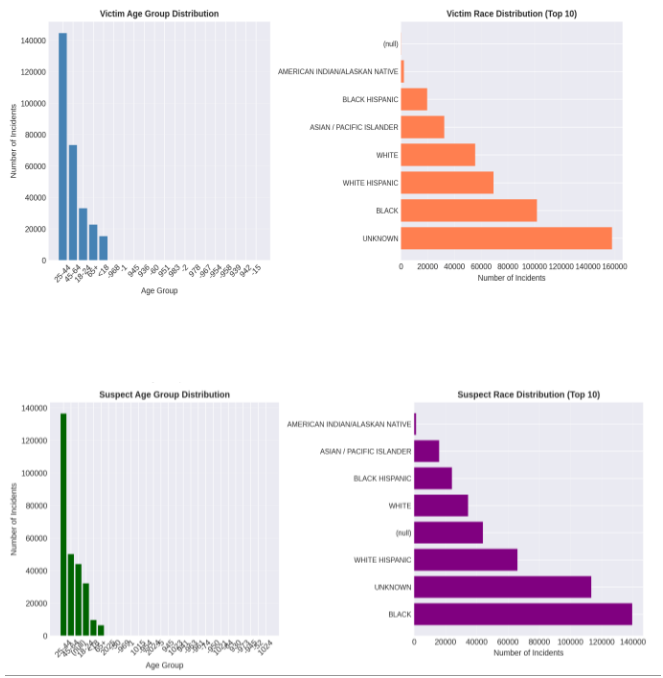

**Figure 8: Demographic Characteristics in Crime Data.**

Distribution of age, race, and sex for both victims and suspects. These patterns reflect broader societal issues and the demographic composition of NYC neighborhoods.

The demographic analysis reveals important patterns in both victim and suspect characteristics:

- Age distribution shows concentration in the 25-44 age group
- Gender patterns vary significantly by crime type
- Racial composition reflects the diversity of NYC's population
- Missing demographic data is more prevalent for suspect information, indicating challenges in witness identification and case resolution.

### 4.3 Advanced Analytical Results

*4.3.1 Correlation Analysis*

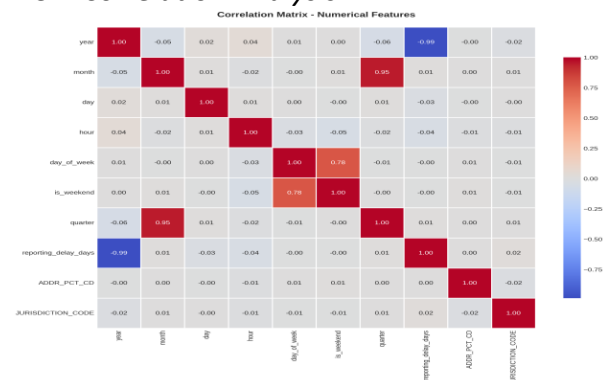

**Figure 9: Comprehensive Correlation Analysis.**

Pearson and Spearman correlation matrices revealed the relationships between temporal, spatial, and categorical variables. Strong correlations inform multivariate modeling approaches[12].

*4.3.2 Outlier Detection*

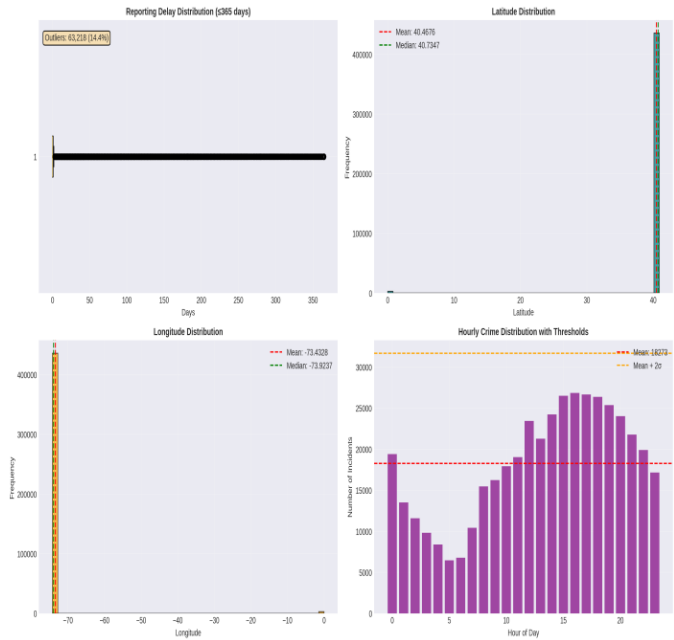

**Figure 10: Statistical Outlier Detection in Crime Data.**
Box plots and distribution analyses identified anomalous observations that may represent unusual crime events, data entry errors, or exceptional circumstances requiring investigation[13].

*4.3.3 Time Series Dependencies*

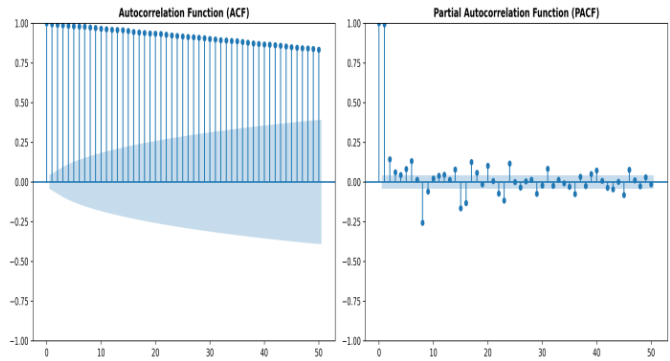

**Figure 12: Time Series Analysis of Crime Patterns.**
The ACF and PACF plots reveal temporal and seasonal patterns in the data, which are essential for forecasting models. Significant autocorrelation at multiple lags indicates a predictable temporal structure[14].

*4.3.4 Dimensionality Reduction*

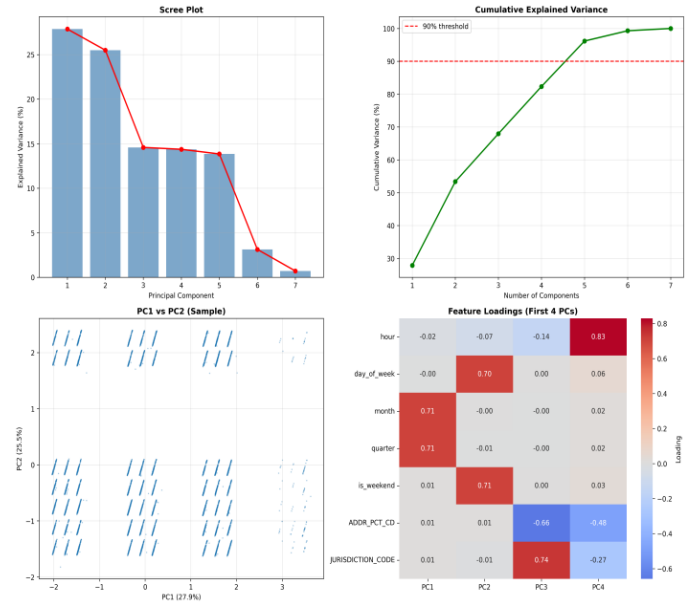

**Figure 14: PCA Variance Decomposition.**
Principal component analysis identifies the key dimensions of variance in the dataset, reducing complexity while retaining critical information. The scree plot and component loadings guide the feature selection for the predictive models[15].

**4.4 Statistical Hypothesis Test Results**

All four statistical tests yielded highly significant results, providing strong evidence against all the null hypotheses.

*4.4.1 Crime Type vs Borough Association*

**Chi-Square Test Results:**
- **Test Statistic**: $\chi^2$ = 8,380.98
- **P-value**: < 0.001
- **Conclusion**: **Statistically Significant**

The extremely high chi-square statistic and infinitesimal p-value indicated a strong association between crime type and borough location. Different boroughs exhibit distinct crime-type profiles, suggesting that geographic context substantially influences the nature of criminal activity. This finding supports targeted location-specific crime prevention strategies.

*4.4.2 Crime Distribution Across Boroughs*

**Kruskal-Wallis H-Test Results:**
- **Test Statistic**: H = 132.59
- **P-value**: $1.09 \times 10^{-27}$
- **Conclusion**: **Statistically Significant**

The Kruskal-Wallis test confirmed that crime frequency distributions differed significantly across boroughs. An exceptionally low p-value indicates that at least one borough's crime pattern differs markedly from that of the others. This validates Brooklyn's identification as the highest-crime borough and suggests heterogeneous crime landscapes across NYC's geographic divisions.

*4.4.3 Weekend vs Weekday Crime Patterns*

**Mann-Whitney U Test Results:**
- **Test Statistic**: U = 15,532,347,295
- **P-value**: $7.88 \times 10^{-111}$
- **Conclusion**: **Statistically Significant**

The Mann-Whitney U test revealed profound differences between the weekend and weekday crime patterns. The astronomically high U-statistic and virtually zero p-value demonstrate that temporal weekly rhythms significantly influence criminal activity. This finding has important implications for police staff and for patrol scheduling.

*4.4.4 Crime Severity and Time of Day*

**Chi-Square Test Results:**
- **Test Statistic**: $\chi^2$ = 2,925.31
- **P-value**: < 0.001
- **Conclusion**: **Statistically Significant**

The chi-square test confirms that crime severity (felony/misdemeanor/violation) varies significantly with time of day. Combined with this result, the peak hour finding (4:00 PM) suggests that both crime volume and severity exhibit temporal patterns, enabling time-sensitive resource-allocation strategies.

## 4.5 Summary of Findings

All statistical tests rejected their respective null hypotheses at a significance level of α = 0.05, with p-values far below conventional thresholds. This comprehensive statistical validation provides robust evidence for the following:
1. Geographic heterogeneity in crime patterns across NYC boroughs
2. Distinct crime type distributions associated with specific locations
3. Temporal variations in criminal activity between weekends and weekdays
4. Time-of-day effects on both crime volume and severity

## 5. Discussion

## 5.1 Interpretation of Spatial Patterns

Brooklyn's emergence as the highest crime borough may reflect multiple factors including population density, socioeconomic diversity, and urban infrastructure characteristics. The statistically significant association between crime type and borough suggests that each borough faces unique public safety challenges that require tailored intervention.

The identification of 78 distinct precincts provides granularity for micro-level analysis, enabling police departments to implement hyperlocal strategies based on precinct-specific patterns.

## 5.2 Temporal Dynamics

The peak crime hour at 4:00 PM aligns with theories of routine activities and urban rhythms. The late afternoon represents a transitional period when
- Workdays conclude and commuting activity peaks
- School dismissals increase juvenile presence in public spaces
- Commercial establishments experience high customer traffic
- Reduced natural surveillance may occur in some areas

Friday identification as the peak crime day corresponds to well-documented weekend patterns. The statistically significant difference between weekend and weekday crime patterns (Mann-Whitney U test) validates the importance of temporal considerations in resource allocation.

## 5.3 Crime Type and Severity Insights

Petit larceny's dominance as the most common crime type reflects broader national trends in property crimes. The 32.38% felony rate indicates that while property crimes are numerically dominant, serious violence and property felonies constitute a substantial proportion of the total complaints.

A 98.75% completion rate suggests that most criminal incidents progressed beyond the attempt stage. This high rate may indicate the following.
- Effective criminal tactics

- Limited prevention or intervention during incidents
- Reporting biases favoring completed over attempted crimes

**5.4 Implications for Law Enforcement**

The statistically validated findings support several evidence-based policing strategies:
1. **Geographic Resource Allocation**: Deploy additional resources to high-crime boroughs, particularly Brooklyn, considering borough-specific crime-type profiles.
2. **Temporal Staffing Optimization**: Increase in patrol presence during peak hours (afternoon or evening) and peak days (Friday) with differentiated weekend strategies
3. **Crime-Specific Prevention**: Develop targeted interventions for petit larceny, including retail security partnerships and public awareness campaigns.
4. **Predictive Policing**: Identify identified spatiotemporal patterns as inputs for predictive models to anticipate crime hotspots

**5.5 Limitations**

Several limitations warrant consideration:
1. **Reporting Bias**: Complaint data reflect only reported crimes, and unreported incidents remained invisible in the analysis.
2. **Temporal Coverage**: While 47-year span provides a longitudinal perspective, structural changes in urban environments and policing practices may complicate trend interpretation.
3. **Causality**: Statistical associations do not establish causal mechanisms, and additional research is required to identify the underlying drivers.
4. **Data Quality**: Missing values in some fields (demographic information and precise locations) limit certain analytical dimensions.

## 6. Conclusions

This comprehensive statistical analysis of NYPD complaint data spanning 47 years provides empirical evidence of significant spatial and temporal patterns in urban crime. The key contributions of this study are as follows:
1. **Quantitative Validation**: All four statistical hypothesis tests yielded highly significant results ($p < 0.001$), providing robust evidence for crime pattern heterogeneity across geographic, temporal, and categorical dimensions.
2. **Actionable Insights**: Identification of Brooklyn as the highest-crime borough, 4:00 PM as peak crime hour, Friday as peak crime day, and Petit Larceny as the most common offense provides concrete targets for intervention.
3. **Methodological Rigor**: Application of multiple statistical testing approaches (Chi-Square, Kruskal-Wallis, Mann-Whitney U) ensures that the findings are not artifacts of single analytical methods.
4. **Policy Relevance**: Results directly inform evidence-based policing strategies, including resource allocation, patrol scheduling, and crime prevention program design.

**6.1 Future Research Directions**

Future investigations should explore:
1. **Longitudinal Trend Analysis**: Examine how crime patterns have evolved across the 47-year period to identify long-term trends and policy impacts.
2. **Multivariate Modeling**: Develop predictive models incorporating multiple variables simultaneously to forecast crime risk
3. **Socioeconomic Integration**: Link crime patterns with demographic, economic, and environmental variables to identify the root causes.
4. **Intervention Evaluation**: Assess the effectiveness of policies implemented based on these findings through quasi-experimental designs.
5. **Real-Time Analytics**: Develop systems for continuous monitoring and near-real-time pattern detection to enable rapid response

**6.2 Final Remarks**

This study demonstrated the power of rigorous statistical analysis applied to large-scale administrative data to generate actionable insights into public safety. The convergence of evidence across multiple statistical tests provides confidence in the identified patterns and supports their use in operational decision making. As cities worldwide face evolving public safety challenges, the data-driven approaches exemplified in this study offer promising pathways toward more effective, efficient, and equitable law enforcement practices.

The 47-year temporal span of this dataset represents a remarkable resource for understanding urban crime dynamics, and continued analysis of NYPD complaint data will undoubtedly yield additional insights as new analytical methods emerge and additional years of data accumulate.

## Acknowledgments

We acknowledge the New York City Police Department for maintaining comprehensive crime records and making them available for research purposes through the NYC Open Data Initiative. We also recognize the broader Open Data movement to promote transparency and evidence-based policy development.

Appendix A: Summary Statistics Table

| Category | Metric | Value |
|---|---|---|
| Temporal | Total Years | 47 |
| | Date Range Start | 1963-05-01 |
| | Date Range End | 2025-09-30 |
| | Peak Hour | 16 (4:00 PM) |
| | Peak Day | Friday |
| Spatial | Total Boroughs | 6 |
| | Total Precincts | 78 |
| | Most Crime Borough | Brooklyn |
| Crime | Total Crime Types | 57 |
| | Most Common Crime | Petit Larceny |
| | Felony Percentage | 32.38% |
| | Completion Rate | 98.75% |
| Dataset | Total Records | 438,556 |

Appendix B: Statistical Test Results Table

| Test | Test Statistic | P-value | Significance |
|---|---|---|---|
| Chi-Square (Crime Type vs Borough) | 8,380.98 | < 0.001 | Yes |
| Kruskal-Wallis (Crime by Borough) | 132.59 | $1.09 \times 10^{-27}$ | Yes |
| Mann-Whitney U (Weekend vs Weekday) | 15,532,347,295 | $7.88 \times 10^{-111}$ | Yes |
| Chi-Square (Severity vs Time of Day) | 2,925.31 | < 0.001 | Yes |